\documentclass[twocolumn,superscriptaddress,preprintnumbers,amsmath,amssymb]{revtex4}

\usepackage{graphicx}

\makeatletter
\def\@dotsep{4.5}
\makeatother

\begin{document}
\title{Real-time observation of oxidation and photo-oxidation of rubrene thin
  films by spectroscopic ellipsometry}

\author{M.~Kytka}

\affiliation{Institut f\"ur Angewandte Physik, Universit\"at T\"ubingen, Auf
  der Morgenstelle 10, 72076 T\"ubingen, Germany}

\affiliation{Faculty of Electrical Engineering and Information Technology,
  Slovak University of Technology in Bratislava, Ilkovi\v cova 3, 812 19
  Bratislava, Slovak Republic}

\author{A.~Gerlach}

\affiliation{Institut f\"ur Angewandte Physik, Universit\"at T\"ubingen, Auf
  der Morgenstelle 10, 72076 T\"ubingen, Germany}

\author{J.~Kov\'a\v c}

\affiliation{Faculty of Electrical Engineering and Information Technology,
  Slovak University of Technology in Bratislava, Ilkovi\v cova 3, 812 19
  Bratislava, Slovak Republic}

\author{F.~Schreiber}
\email{frank.schreiber@uni-tuebingen.de}

\affiliation{Institut f\"ur Angewandte Physik, Universit\"at T\"ubingen, Auf
  der Morgenstelle 10, 72076 T\"ubingen, Germany}

\date{\today}

%
%

\begin{abstract}
  We follow in real-time and under controlled conditions the oxidation of the
  organic semiconductor rubrene grown on SiO$_2$ using spectroscopic
  ellipsometry.  We derive the complex dielectric function $\varepsilon_1$ +
  i$\varepsilon_2$ for pristine and oxidized rubrene showing that the oxidation
  is accompanied by a significant change of the optical properties, namely the
  absorption.
  We observe that photo-oxidation of rubrene is orders of magnitude faster
  than oxidation without illumination.
  By following different absorption bands (around $2.5\,$eV and $4.0\,$eV for
  pristine rubrene and around $4.9\,$eV for oxidized rubrene) we infer that
  the observed photo-oxidation of these films involves non-Fickian diffusion
  mechanisms.
\end{abstract}

\maketitle

\begin{figure}[htbp]
\centering
\includegraphics[width=85mm]{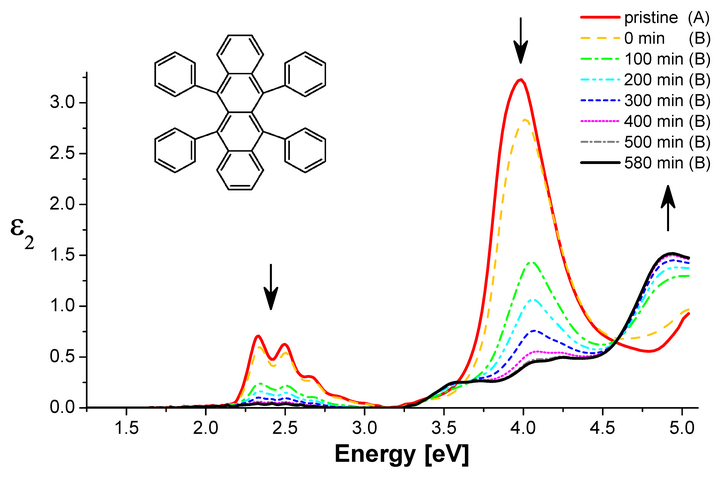}
\caption{Dielectric function $\varepsilon _2$ of the
  rubrene film before and during photo-oxidation (initial film thickness
  $25.1\,$nm). The absorption bands around $2.5\,$eV and $4.0\,$eV correspond
  to pristine rubrene, whereas the feature at $4.95\,$eV originates from
  oxidized rubrene.  The relatively small difference between the pristine and
  '$0\,$min' spectrum shows effect of oxidation after $870\,$min without
  illuminating the sample.}
\label{fig:Absorption}
\end{figure}

Many organic materials with delocalized $\pi$-electron systems exhibit
significant potential for electronic and optoelectronic applications
\cite{Schreiber_2004_PSS-A_201_1037}. But despite progress in the development
of encapsulation strategies \cite{Sellner_2004_AM_16_1750} one of the
important issues in this area remains the change of the electronic properties
upon exposure to ambient gases.  Rubrene (C$_{42}$H$_{28}$,
5,6,11,12-tetraphenylnaphthacene, see inset of Fig.~\ref{fig:Absorption})
belongs to a group of small organic molecules with promising properties which
found use in organic light emitting diodes (as a red dopant)
\cite{Aziz_2002_APL_80_2180} and organic field effect transistors
\cite{Boer_2004_PSS-A_201_1302}.  However, as also other molecules
\cite{Vollmer_2006_SS_600_4004}, rubrene exhibits strong reactivity and
affinity to (photo-)oxidation which reduces the stability and lifetime of
devices.  While it is well known that rubrene tends to undergo
(photo-)oxidation \cite{Hochstrasser_1956}, the understanding and control of
degradation due to oxidation of these materials still is a key challenge in
organic electronics.  For this purpose we studied in real-time and under
controlled conditions the kinetics of oxidation of rubrene thin films using
spectroscopic ellipsometry.

\vspace{3mm}
The rubrene material used was purchased from Acros and purified by gradient
sublimation.  The Si(100) substrates with $\sim 20\,$nm thermal oxide were
cleaned with acetone and propanol in an ultrasonic bath, transfered into the
vacuum chamber (base pressure $p = 3 \times 10^{-8}\,$mbar) and heated at
temperatures $T \geq 400 \,^\circ$C for several hours. The rubrene films were
grown with the substrate at room temperature by evaporation from a Knudsen
cell with a typical growth rate of $0.85\,$nm/min. The experimental data were
acquired \textit{in situ} using a spectroscopic ellipsometer (Woollam
\mbox{M-2000}) with a broad band $75\,$W Xe-lamp ($250\,$nm to $1000\,$nm) and
CCD-based detection system with a resolution of about $1.6\,$nm. The light
spot on the sample was $\sim 2\times 6\,$mm$^2$.  The light source and
detector were mounted to the vacuum chamber which provides a pair of
strain-free windows at a relative angle of $120^\circ$.

After growing the film in high vacuum the oxidation experiment was performed
in the same chamber.  For this purpose we admitted $40\,$mbar air using a leak
valve. To distinguish oxidation and photo-oxidation effects we studied two
separate spots on the sample. The first spot (A) was investigated in high
vacuum and yields the spectrum of the pristine rubrene. After $870\,$min
exposure to air without illumination with the Xe-lamp we switched to a
different spot (B) which was illuminated continuously during the
photo-oxidation.  Ellipsometry scans with $100$ compensator revolutions were
taken every $0.2\,$min at the beginning and every $2\,$minutes later.

Using a commercial software package (WVASE32) we converted the ellipsometric
data to the dielectric function $\varepsilon_1 + i \varepsilon_2$.  Since
rubrene grows in non-crystalline films
\cite{Kowarik_2006_PChChP_8_1834,Kafer_2005_PRL_95_166602}, we employed an
isotropic 4-layer model: rubrene -- SiO$_2$ -- Si$_\mathit{int}$ -- Si, with
the optical constants of Si, SiO$_2$ and Si$_\mathit{int}$ (interfacial layer
between SiO$_2$ and Si) taken from Ref.~\cite{Herzinger_1998_JoAP_83_3323}.
Inhomogeneities of the organic layer during oxidation are neglected in this
effective medium approximation. The essential changes of the spectra upon
exposure to air, however, were found to prevail regardless of the
ellipsometric model used for the analysis. To cross-check our results we
verified that both the spectrum of the pristine and the fully oxidized film
agree with earlier studies on rubrene \cite{Otomo_2002_OL_27_891}.
Starting with the spectrum of the bare substrate we determined the thickness
of the SiO$_2$ and Si$_\mathit{int}$ layer as well as the precise angle of
incidence. The optical properties of the organic film were obtained by
point-by-point fits, the film thickness by employing a Cauchy model in the
transparent region. Consistency checks for pristine rubrene with a general
oscillator model gave good agreement over the complete spectral range.  A more
detailed discussion of the fitting procedure and the optical properties of
pristine rubrene will be published elsewhere \cite{Kytka}.


As shown in Fig.~\ref{fig:Absorption} the optical properties change
significantly throughout the oxidation process. Since the organic film becomes
more and more transparent, even visual inspection reveals the strong effect of
oxidation.  The spectrum of the pristine rubrene and the spectrum taken after
$870\,$min on the second spot (denoted as '$0\,$min' illumination in
Fig.~\ref{fig:Absorption}) do not differ strongly. Hence we conclude that
oxidation without illumination is relatively slow compared to the subsequent
photo-oxidation. While the absorption bands around $E= 2.5$ and $4\,$eV
decrease rapidly during oxidation, a feature centered at $E = 4.95\,$eV
appears. We further note that the spectra shown in Fig.~\ref{fig:Absorption}
exhibit an isosbestic point at $4.59\,$eV, where the absorption
$\varepsilon_2^\mathit{iso}$ remains constant during oxidation.  This
indicates that the organic film can essentially be regarded as a non-volatile
two component system, i.e.\ pristine and oxidized rubrene, without noticeable
intermediate or products.

\begin{figure}[htbp]
  \centering 
 \includegraphics[width=85mm]{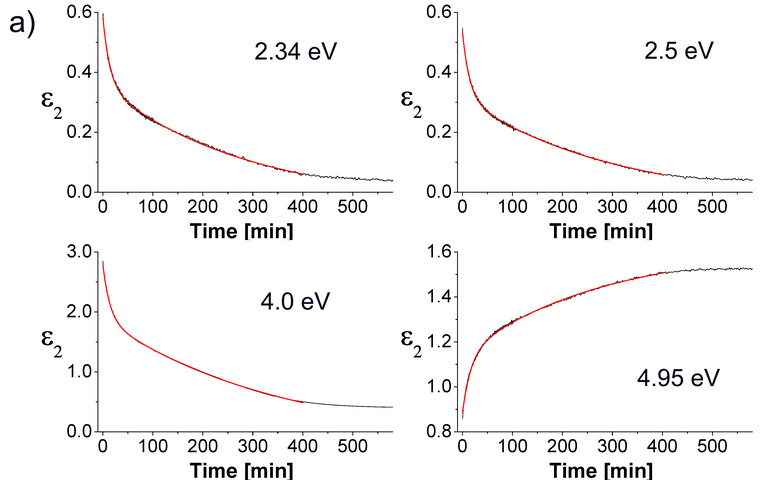}
 \includegraphics[width=85mm]{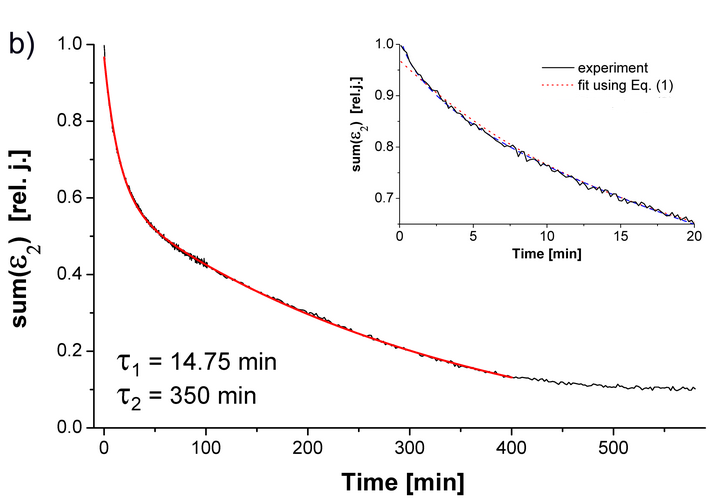}
\caption{Time dependence of the dielectric constant with a fit according
  to Eq.~(\ref{eq:2ExpDec}) a) for selected energies in the spectra, i.e.\ 
  $2.34\,$eV ($0-0$ transition), $2.5\,$eV ($0-1$ transition), $4.0\,$eV
  (strongest absorption feature), and $4.95\,$eV (absorption band of oxidized
  rubrene).  b) integral of $\varepsilon _2$ over the interval
  $1.25-3.17\,$eV.}
\label{fig:Epsilon_sum}
\end{figure}

Figure~\ref{fig:Epsilon_sum}a illustrates the time dependence of the
dielectric function for selected energies, i.e.\ $2.34$, $2.5$, $4.0$, and
$4.95\,$eV, whereas Fig.~\ref{fig:Epsilon_sum}b shows the integral of
$\varepsilon _2$ over the interval $1.25-3.17\,$eV.  The decrease of the
absorption bands around $2.5\,$ and $4.0\,$eV (pristine rubrene) goes hand in
hand with the increase of the high energy feature around $4.95\,$eV (oxidized
rubrene).  Apparently, the time dependence of the dielectric constants is
non-trivial.  While initially the oxidation is very rapid, it is slowing down
after several minutes. Empirically, we find that $\varepsilon_2(t)$ can be
described by the sum of two exponentials, i.e.\ 
\begin{equation}
  y(t) = y_0 + \alpha_1 \, \exp(-t/\tau_1) + \alpha_2 \, \exp(-t/\tau_2).
  \label{eq:2ExpDec}
\end{equation}
Although the absolute values of $\tau_1$ and $\tau_2$ depend slightly on the
given sample and experimental conditions, we invariably find two very
different time constants (see Fig.~\ref{fig:Epsilon_sum}b), i.e.\ $\tau_1 /
\tau_2 \ll 1$, indicating that two different mechanisms are involved.

\begin{figure}[htbp]
\centering
\includegraphics[width=85mm]{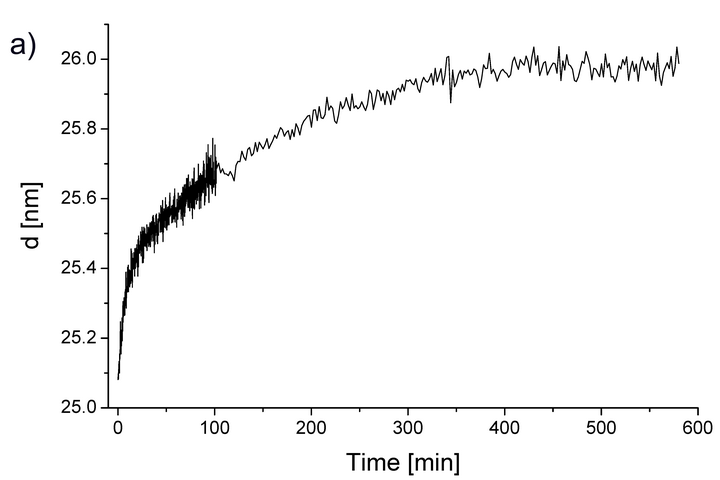}
\includegraphics[width=85mm]{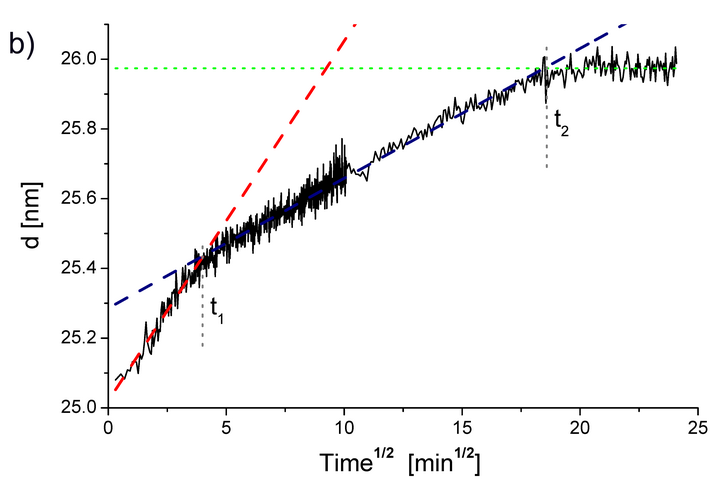}

\caption{Thickness evolution $d(t)$ of the rubrene film during
  photo-oxidation. (a) The real-time data derived from ellipsometry
  measurements exhibit two regimes.  (b) The data on a $t^{1/2}$-axis with a
  fit according to a non-Fickian model, see Eq.~(\ref{eq:Sqrt}).}
\label{fig:Thickness}
\end{figure}

The film thickness $d(t)$ derived from the ellipsometry data exhibits two
regimes, see Fig.~\ref{fig:Thickness}. Starting at $d_0 = 25.1\,$nm we find a
rapid expansion at the beginning and a slowly saturating behavior after an
thickness increase of $0.9\,$nm towards the end of oxidation.
The similarity with the time dependent absorption at $4.95\,$eV, i.e.\ a
feature related to the oxidized rubrene, suggests that the thickness increase
is caused by the intake of oxygen. Provided that each rubrene molecule reacts
with one oxygen molecule to rubrene peroxide \cite{Kafer_2005_PCCP_7_2850} a
volume expansion of 2.1\% is expected (assuming a constant density of the
organic).  For fixed lateral dimensions the thickness would consequently
increase by a maximum of 6.5\% --- an estimate which is compatible with the
observed rise by 4.2\% and which shows that nearly the entire film can be
oxidized.

To explore the kinetics of rubrene oxidation in more detail we tried several
models describing the data.  Interestingly, within the given dynamic range a
good fit to the thickness evolution cannot only be obtained with an
exponential expression as in Eq.~(\ref{eq:2ExpDec}) but also with a composite
square-root dependence, i.e.\ 
\begin{equation}
  d(t)  \ = \left\{ 
    \begin{array}{ll} 
      d_0 + a t^{1/2} & \mathrm{for \ } 0 \leq t  \leq t_1 \\ 
      d_0' + b t^{1/2} & \mathrm{for \ } t_1 < t \leq t_2 \\
      \mathrm{const.} & \mathrm{for \ } t > t_2
  \end{array}\right. ,
  \label{eq:Sqrt}
\end{equation}
with $t_1 \approx 15\,$min ($t_1^{1/2} \approx 3.87\,$min) and $t_2 \approx
360\,$min ($t_1^{1/2} \approx 18.97\,$min) for the data presented in
Fig.~\ref{fig:Thickness}b.  Thus the oxidation kinetics do not follow a simple
Fickian diffusion model which would result in a continuous
$t^{1/2}$-dependence.  The underlying reason for this behavior cannot easily
be identified, however, there are several possible explanations for this
non-Fickian diffusion: for example
(1) a depth dependent diffusivity of oxygen,
(2) space charge effects,
(3) surface effects influencing the intake of oxygen, or
(4) a delay between the diffusion and the oxidation reaction.
In order to distinguish these effects and their influence on the oxidation
kinetics of rubrene further studies are required. Complementary techniques and
a theoretical description could improve our understanding of the described
phenomenon considerably.

In conclusion we have presented real-time ellipsometry data which yield
essential information about the optical properties and the reaction kinetics
of rubrene.  Based on the thickness data we infer that the diffusion of oxygen
into the organic follows a non-Fickian law.  We hope our study on the
oxidation of rubrene will contribute to an improved understanding of
degradation mechanisms in organic electronic materials.

\section*{Acknowledgments}

The authors thank J.\ Pflaum for providing the purified rubrene and G.\ Witte
for discussing the results of our work.  We gratefully acknowledge financial
support by the DFG and EPSRC.

\end{document}